\documentclass[twoside]{article}
\usepackage{fleqn,espcrc2,psfig}

\usepackage[figuresright]{rotating}

\def\pixdir{.}

\newcommand{\AmS}{{\protect\the\textfont2
  A\kern-.1667em\lower.5ex\hbox{M}\kern-.125emS}}

\title{
\vspace*{-0.6in}
\hspace*{5in}
{\normalsize UM-P-044-2000}\\
\hspace*{5in}
{\normalsize RCHEP-010-2000}\\
\vspace*{0.3in}
Pad\'e--Z$_2$ stochastic estimator of determinants applied to quark loop 
expansion of lattice QCD.
}

\author	{
	J.F. Markham
	\address
		{
		School of Physics, 
		University of Melbourne, 
		Vic 3052, Australia
		}
       	T. D. Kieu
	\address 
		{
		CSIRO MST, 
		Private Bag 33, 
		Clayton South MDC,
		Vic 3169, Australia
		}
	}
       
\begin{document}

\begin{abstract}
We use the Pad\'e--Z$_2$ 
stochastic estimator for calculating 
$\mbox{Tr}\log{\bf M}$ to compute
quark loop corrections to quenched QCD.
We examine the main source of error in this technique and look at a way of controlling it.
\vspace{1pc}
\end{abstract}

\maketitle

\section{Introduction}
In lattice QCD calculation, evaluating the fermion determinant is 
computationally expensive
and as a consequence $\det{\bf M}$ is often not evaluated and 
is simply set equal to a constant.
This is called the {\em quenched approximation} 
and amounts to neglecting internal fermion loops.
The goal of this work is to study the systematic error introduced by
quenching and to
compensate for quenching errors by constructing
an expansion in quark-loop count 
and using it to improve Monte Carlo estimators of Wilson loops
measured on quenched gauge field configurations.
We examine a source of statistical error and a method for its
partial alleviation.

\section{Re-weighting existing quenched QCD configurations}
With the full QCD action
\begin{equation}
S[U,\psi,\bar\psi] = S^{Wilson}_{fermion}[U] + S_{gauge}[U,\psi,\bar\psi] \;,
\label{eq:QCD_lattice_action}
\end{equation}
the expectation value of some operator, ${\cal O}$ is given by
\begin{eqnarray}
\left\langle {\cal O}\right\rangle  &=& \frac{1}{Z} \int[dU][d\psi][d\bar\psi] {\cal O}e^{-S[U,\psi,\bar\psi]} \nonumber \\
&=& \frac{1}{Z} \int[dU] {\cal O}det{\bf M}[U]e^{-S_{gauge}[U]} \;,
\label{eq:O_expectation}
\end{eqnarray}
where 
\begin{eqnarray}
Z &=& \int[dU] det{\bf M}[U]e^{-S_{gauge}[U]} \;.
\label{eq:Z_QCD}
\end{eqnarray}

By rewriting $S^{Wilson}_{fermion}[U]$ as
\begin{eqnarray}
-S^{Wilson}_{fermion}[U] &=& n_f \log{\det{{\bf M}[U]}} \nonumber\\
&=&	 n_f \mbox{Tr}\log{\bf M}[U] \nonumber\\ 
&=& \delta S[U] \;,
\label{eq:delta_s_log}
\end{eqnarray}
and dividing through by  $e^{\left\langle \delta S\right\rangle }$, (\ref{eq:O_expectation}) becomes
\begin{eqnarray}
\left\langle {\cal O}\right\rangle  &=& \frac{\int[dU] {\cal O}e^{-S_{\rm gauge}[U]} e^{(\delta S[U]-\left\langle \delta S\right\rangle )} }
{\int[dU] e^{-S_{\rm gauge}[U]} e^{(\delta S[U]-\left\langle \delta S\right\rangle )}} \;.
\label{eq:O_reweighted}
\end{eqnarray}
Expanding the exponential in the fluctuations in $\delta S$ leads to
\begin{eqnarray}
\left\langle {\cal O}\right\rangle  
&=& \left\langle {\cal O}\right\rangle _{_{\delta S = 0}} \nonumber \\
&& + \left\langle {\cal O} \delta S\right\rangle _{_{\delta S = 0}} -
\left\langle {\cal O}\right\rangle _{_{\delta S = 0}} \left\langle \delta S\right\rangle _{_{\delta S = 0}} \nonumber \\
&& + \cdots \;\;.
\label{eq:O_reweighted_first_order}
\end{eqnarray}
Here, the subscript on $\left\langle {\cal O}\right\rangle _{_{\delta S = 0}}$ signifies the quenched
expectation value of ${\cal O}$.

\section {Pad\'e--Z$_2$ stochastic estimator for evaluating $\mbox{Tr}\log {\bf M}$}
In order to make use of (\ref{eq:O_reweighted_first_order})
one needs to be able to find $\delta S$.
We use the 
Pad\'e--Z$_2$ method to find $\mbox{Tr}\log {\bf M}$ as per \cite{Thron:1997iy}.

The Pad\'e approximant for $\mbox{Tr}\log {\bf M}$ can be written as 
\begin{equation}    
\mbox{Tr}\log {\bf M} \approx b_0 Tr {\bf I} +
               \sum_{k=1}^K  b_k\cdot \mbox{Tr}({\bf M} +c_k {\bf I})^{-1}.
\end{equation}

$\mbox{Tr} ({\bf M} +c_k {\bf I})^{-1}$ can be estimated using noise Z$_2$ vectors, $\eta^j \;$.
\begin{eqnarray}
\mbox{Tr}({\bf M}+c_k)^{-1} &\approx& \frac 1L \sum_{j}^L \eta^{j \dagger} (\xi^{k,j}),
\end{eqnarray}
where $\xi^{k,j}= ({\bf M}+c_k {\bf I})^{-1} \eta^j$ are the solutions of
\begin{eqnarray}
({\bf M} + c_k {\bf I} )\xi^{k,j} &=& \eta^j
\label{eq:original_AX_equals_b}
\end{eqnarray}
where $j=1,\cdots,L$ and $k=1,\ldots,K$.
The variance of these estimators can be greatly reduced by subtracting
suitably chosen traceless matrices
\begin{eqnarray}
\mbox{Tr}({\bf M}+c_k {\bf I})^{-1} &\approx&
< \eta^\dagger (({\bf M}+c_k {\bf I})^{-1} \\\nonumber
&&- \sum_{p=1}^P \lambda_{p,k} {\bf Q}^{(p)})) \eta>
\end{eqnarray}
where the $\lambda_{p,k}$ are chosen to minimise the variance of the estimator. 
The ${\bf Q}^{(p)}$ used are   
\begin{eqnarray}
{\bf Q}^{(p)} &=&\frac {\kappa^p}{(1+c_k)^{p+1}}({\bf D}^p-{\rm Tr}{\bf D}^p)\;, \nonumber
\end{eqnarray}
where ${\bf M} = {\bf 1} - \kappa {\bf D}$. The odd powers of ${\bf D}$ are traceless,
but the even powers greater than two are not. $\mbox{Tr}~{\bf D}^4$ and 
$\mbox{Tr}~{\bf D}^6$ are calculated from $4$ and $6$ link loops respectively.
For this work we expand to order $K=11$, using $L=10$ noise vectors and $P=8$ traceless
subtraction matrices. The variational procedure to set $\lambda_{p,k}$ was found not
to be necessary and so we set them equal to unity. 
To solve (\ref{eq:original_AX_equals_b}) we used the MR$\rm ^3$ algorithm from 
\cite{Glassner:1996gz} and the odd--even preconditioner in from \cite{Frommer:1995ik}.

\section {Application to Wilson loops}
\begin{figure}[htb]
\psfig{figure=\pixdir/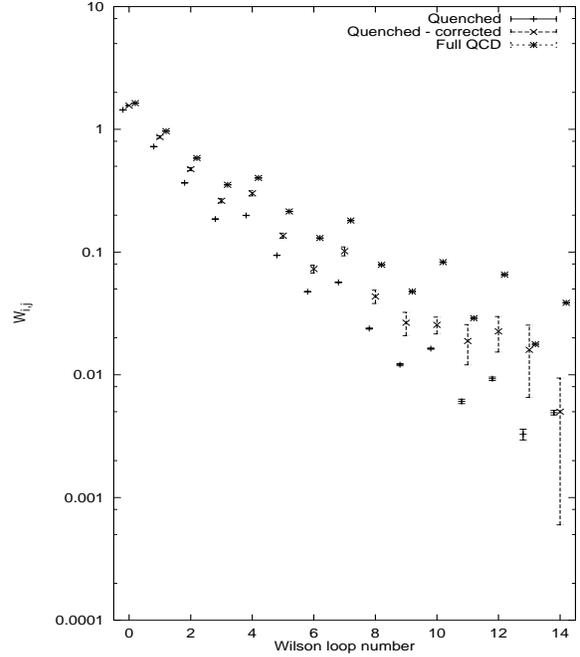,height=3.5in,width=3.0in}
\vspace{-.15in}
\caption{Effect of correcting Wilson loops by expanding to first order in 
$\delta S[U] = n_f \mbox{Tr}\log{\bf M}[U]$ as per (\ref{eq:O_reweighted_first_order}).
}
\label{fig:NoDeltaBetaWilson}
\end{figure}

We apply the preceding numerical techniques to finding improved estimates
of Wilson loops.
Measurements were made on two sets of $10^4$ gauge field configurations;
a set of $103$ quenched QCD configurations and a set of $95$ full QCD configurations, 
generated using the MILC collaboration software \cite{{MILC_freehep}}.

Both have $\beta = 5.436$, and the full QCD configurations have
$\kappa = 0.16$ and  $n_f = 2$.
Fig.~\ref{fig:NoDeltaBetaWilson} shows the results for various Wilson loops
with corrections done to first order as per (\ref{eq:O_reweighted_first_order})
In all of Figs.~\ref{fig:NoDeltaBetaWilson} -- \ref{fig:YesDeltaBetaWilsonRelativeLimitedSum}
the loop number on the x--axis labels $W_{ij}$ from left to right in the following order:
$W_{_{1\,1}}$, $W_{_{1\,2}}$, $W_{_{1\,3}}$, $W_{_{1\,4}}$, $W_{_{1\,5}}$, $W_{_{2\,3}}$, $W_{_{1\,7}}$, 
$W_{_{2\,4}}$, $W_{_{1\,9}}$, $W_{_{3\,3}}$, $W_{_{1\,10}}$, $W_{_{2\,5}}$ .
In Figs.~\ref{fig:NoDeltaBetaWilson},~\ref{fig:NoDeltaBetaWilsonRelative} and 
\ref{fig:YesDeltaBetaWilsonRelative}, points have been offset slightly for clarity.

\section {A method for controlling the statistical error \label{section:controlling_error} }
\begin{figure}[htb]
\psfig{figure=\pixdir/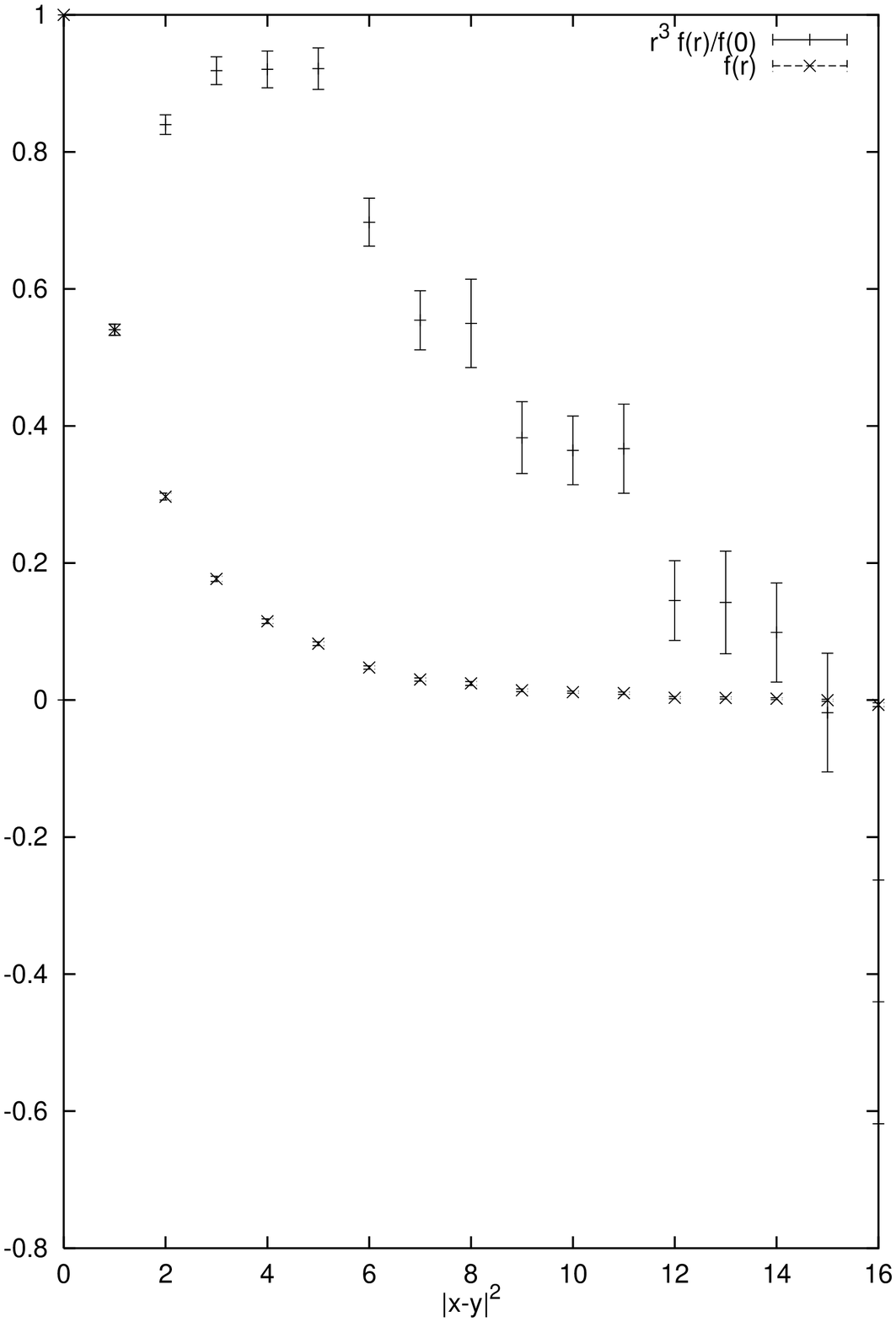,height=3.5in,width=3.0in}
\vspace{-.15in}
\caption{Contribution to $\left\langle \delta {\cal O} \right\rangle $ 
of correlator between 
$(\delta S[U])_{y} = (n_f \log{\bf M}[U])_{yy}$
and $(W_{_{1\,1}})_x$ to 
$\left\langle \delta{\cal O} \right\rangle $}
\label{fig:NoDeltaBetaWilsonLogCorR3}
\end{figure}

\begin{figure}[htb]
\psfig{figure=\pixdir/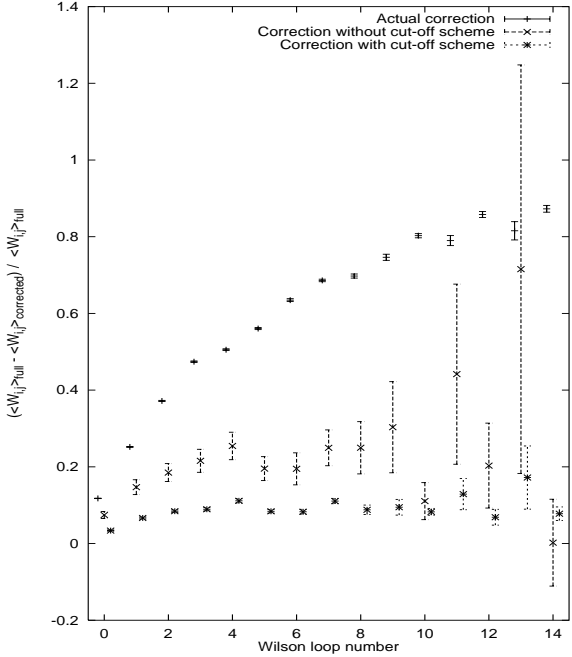,height=3.5in,width=3.0in}
\vspace{-.15in}
\caption{Corrections to Wilson loops relative to the full QCD value
comparing the effectiveness 
with and without the cut off scheme.
}
\label{fig:NoDeltaBetaWilsonRelative}
\end{figure}

A method for controlling the statistical error in (\ref{eq:O_reweighted_first_order})
is suggested by looking at the form of the correlator between the diagonal elements of 
$\log{{\bf M}}$ and the
operator, ${\cal O}$, being measured. Given that ${\cal O}$ and $\delta S$
can be written
\begin{equation}
{\cal O} = \sum_{x} {\cal O}_x \;\;,\;\; \delta S = \sum_{y} (\delta S)_{y}
\end{equation}

then the first order correction to ${\cal O}$ can be written as 
\begin{eqnarray}
\delta \left\langle {\cal O}\right\rangle  &\approx&  
\left\langle \sum_{x}{\cal O}_{x} \sum_{y} \delta S_{y}\right\rangle \nonumber \\
&&- \left\langle \sum_{x}{\cal O}_{x} \right\rangle
\left\langle \sum_{y} \delta S_{y} \right\rangle \nonumber \\
 &=& \sum_{x,y} \left\langle {\cal O}_{x} (\delta S_{y} - \left\langle \delta S_{y}\right\rangle )\right\rangle   \nonumber \\
 &=& \sum_{x,y} \left\langle {\cal O}_{x} \delta \tilde{S}_{y}\right\rangle  
\label{eq:O_reweighted_first_order_per_site}
\end{eqnarray}
If we write $f(r) = \left\langle {\cal O}_x \delta \tilde{S}_{y}\right\rangle $,
where $r = |x-y|$, then
\begin{eqnarray}
\sum_{x,y} \left\langle {\cal O}_{x} \delta \tilde{S}_{y}\right\rangle 
&\rightarrow& \int dr d\Omega r^{3}f(r)
\label{eq:O_reweighted_first_order_polar}
\end{eqnarray}
Fig.~\ref{fig:NoDeltaBetaWilsonLogCorR3} shows $f(r)/f(0)$ and $r^3f(r)/f(0)$ as a function of $r^2$
for the case ${\cal O} = W_{_{11}}$.
The former shows that the two operators are correlated, as would be expected.
The latter shows that most of the signal for $\delta {\cal O}$ is at small $r$ and
but most of the noise is at large $r$, and that for large lattices this
has the potential to swamp the signal.
One way to address the problem is to cut the integral off at some $r_{max}$.
\begin{eqnarray}
\left\langle \delta {\cal O} \right\rangle  = 
\sum_{|x-y| \atop \leq r_{\rm max}} \left\langle {\cal O}_{x} \delta \tilde{S}_{y}\right\rangle
\label{eq:O_reweighted_first_order_limited_sum}
\end{eqnarray}
This has been done using $r_{max} = 4$, and the results 
are shown in Fig.~\ref{fig:NoDeltaBetaWilsonRelative}.
The cut off scheme reduces the size of the statistical error but also looses some signal.

The Monte Carlo estimator for $\left\langle {\cal O}\right\rangle $ is
\begin{eqnarray}
\left\langle \left\langle \delta {\cal O}\right\rangle \right\rangle  = 
\frac{1}{N} \sum_{n=1}^{N} \sum_{|x-y| \atop \leq r_{\rm max}}{\cal O}_{x}^{n} \delta \tilde{S}_{y}^{n}.
\label{eq:O_reweighted_first_order_limited_sum_estimator}
\end{eqnarray}

An efficient way to do this is with Fourier convolution.
Given a windowing function $W(x-y)$, where
\begin{eqnarray}
W(x-y) = \left\{ 
\begin{array}{ll}
1 & \mbox{if $|x-y| \leq r $} \\
0 & \mbox{ otherwise}
\end{array}
\right.
\label{eq:limited_sum_window}
\end{eqnarray}
then (\ref{eq:O_reweighted_first_order_limited_sum_estimator}) becomes
\begin{eqnarray}
\left\langle \left\langle \delta {\cal O}\right\rangle \right\rangle  
&=& \frac{1}{N} \sum_{n=1}^{N} \sum_{x,y}{\cal O}_{x}^{n} \delta \tilde{S}_{y}^{n}W(x,y) \nonumber \\
&=& \frac{1}{N} \sum_{n=1}^{N} \sum_{x} {\cal O}_{x}^{n} \\
&& \times \left [{\cal F}^{-1} \left\{  {\cal F} \{ \delta \tilde{S}_{y}^{n} \} 
{\cal F} \{ W(x,y)\} \right\} \right]_x  \nonumber
\label{eq:O_reweighted_first_order_limited_sum_estimator_fourier}
\end{eqnarray}
where ${\cal F}$ and ${\cal F}^{-1}$ are defined in the usual way
and implemented as fast Fourier transforms. Using this method 
to compute $\left\langle \delta {\cal O} \right\rangle $ 
also makes it easy to choose a more elaborate windowing function should this be desired.

\section{Approximating full QCD by shifting the gauge coupling {\em and} expanding $\mbox{Tr}\log {\bf M}$}
\begin{figure}[htb]
\psfig{figure=\pixdir/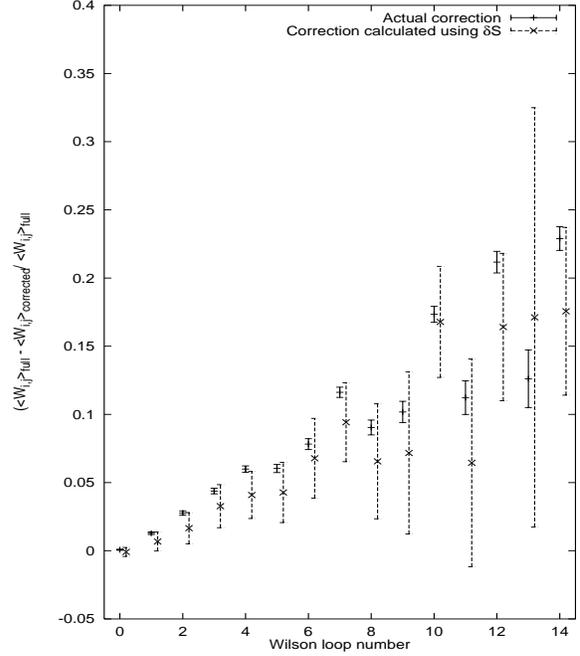,height=3.5in,width=3.0in}
\vspace{-.15in}
\caption{Relative corrections to Wilson loops by expanding to first order in
$\delta S = n_f \mbox{Tr}\log {\bf M} - \delta \beta S_{gauge}$ 
as per (\ref{eq:lw_delta_s})}
\label{fig:YesDeltaBetaWilsonRelative}
\end{figure}

Following \cite{Lee:1998ng} and identifying
\begin{eqnarray}
\delta S = n_f \mbox{Tr}\log {\bf M} - \delta \beta S_{gauge} \;,
\label{eq:lw_delta_s}
\end{eqnarray}
we again look at corrections to Wilson loops.
These are measured on
a set of $200$ $10^4$ full QCD configurations with $\beta = 5.679$,
$\kappa = 0.16$ and  $n_f = 2$ and
also the set of quenched QCD configurations used earlier. 
The results are shown in Fig.~\ref{fig:YesDeltaBetaWilsonRelative} 
and are in accordance with \cite{Lee:1998ng}.
A notable difference between the two is the computational savings afforded by
the choice of algorithm for $\mbox{Tr}\log {\bf M}$.
Use of unbiased subtractors means that only $10$ noise vectors are needed.
In addition to this we do not have to fix the configurations to Landau gauge.

\begin{figure}[htb]
\psfig{figure=\pixdir/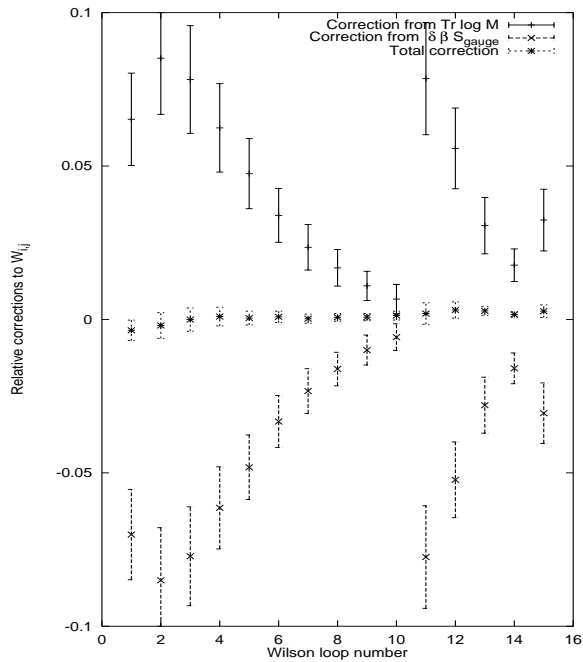,height=3.5in,width=3.0in}
\vspace{-.15in}
\caption{Relative corrections to Wilson loops from each component of $\delta S$ both separately
and in combination.}
\label{fig:YesDeltaBetaWilsonRelativeNoLimitedSum}
\end{figure}

The previously mentioned method of cutting the integral off to reduce the statistical error 
turns out not to work for $\delta S$
defined in (\ref{eq:lw_delta_s}). The contributions from $\mbox{Tr}\log {\bf M}$
and $\delta \beta S_{gauge}$ are nearly equal but are opposite in sign
and so the total correction is small as is shown in 
Fig.~\ref{fig:YesDeltaBetaWilsonRelativeNoLimitedSum}

\begin{figure}[htb]
\psfig{figure=\pixdir/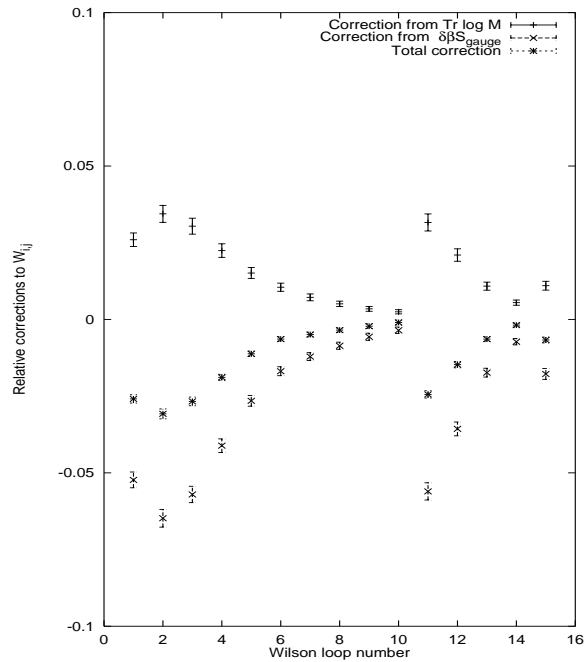,height=3.5in,width=3.0in}
\vspace{-.15in}
\caption{Relative corrections to Wilson loops from each component of $\delta S$ 
using the cut-off scheme to control the statistical error, and with these components combined.}
\label{fig:YesDeltaBetaWilsonRelativeLimitedSum}
\end{figure}

Introducing a cutoff 
reduces the contributions from  $\mbox{Tr}\log {\bf M}$ and $\delta \beta S_{gauge}$
by different amounts which produces a large variation in the total correction.
This is shown in Fig.~\ref{fig:YesDeltaBetaWilsonRelativeLimitedSum}.

We have also looked at corrections to two point functions but the noise is 
such that this method cannot be used {\em as is} to provide improved mass measurements.

\section{Conclusion}
A method of improving estimates of observables measured on quenched QCD configurations
was tested on some sets of small ($10^4$) configurations.
Expanding to first order in $\delta S = n_f \mbox{Tr}\log{\bf M}$ 
gave moderate improvement to estimators for Wilson loops. 
Analysis of the source of the statistical error 
hinted at problems that the method would have on 
larger lattices and with two point functions,
and also suggested a technique
for its partial alleviation.
The technique does have the potential to address these problems
but at the price of reducing the size of the correction.

It was shown in \cite{Lee:1998ng} that combining the above method with an appropriate
shift in the gauge coupling can provide more accurate corrections to 
Wilson loop values. We obtained the same results using a  
more economical method of calculating $\mbox{Tr}\log {\bf M}$.

\section*{Acknowledgements}

We would like to thank Keh-Fei Liu, John Sloan and Don Weingarten for
useful discussions and in particular Keh-Fei Liu for his support of 
John Markham during his visit to the University of Kentucky.

\end{document}